\documentclass[conference]{IEEEtran}
\IEEEoverridecommandlockouts

\usepackage{cite}
\usepackage{amsmath,amssymb,amsfonts}
\usepackage{algorithmic}
\usepackage{graphicx}
\usepackage{textcomp}
\usepackage{xcolor}
\usepackage{hyperref}
\usepackage{multirow}
\usepackage{multicol}
\usepackage{bigstrut}
\usepackage{circledsteps}
\usepackage{subfig}
\usepackage{wrapfig}
\usepackage{algorithm}
\usepackage{soul}
\usepackage[normalem]{ulem}
\usepackage{flushend} % balance two columns on last page

\usepackage[absolute,overlay]{textpos}
\begin{document}
% \pagenumbering{arabic}
\pagestyle{plain}
\title{RayFlex: An Open-Source RTL Implementation of the Hardware Ray Tracer Datapath}

\author{\IEEEauthorblockN{Fangjia Shen, Aaron Barnes, Anusuya Nallathambi, Timothy G. Rogers}
\IEEEauthorblockA{\textit{Elmore Family School of Electrical and Computer Engineering} \\
\textit{Purdue University}\\
West Lafayette, Indiana, U.S.A. \\
\{shen449, barnes88, anallat, timrogers\}@purdue.edu
}
% \and
% \and
% \and
% \and
% \and
}

\maketitle

\begin{abstract}

The advent of hardware ray tracing (RT) units has brought unprecedented realism to real-time rendered computer graphics. However, the potential of these units extends beyond graphics, offering acceleration for various computational tasks such as tree traversal and nearest-neighbor search. We introduce RayFlex, a first-of-its-kind open-source RTL implementation of a hardware ray tracer datapath designed to facilitate research in general-purpose programmable RT units. RayFlex's architecture is both extensible and flexible, thanks to two core design concepts: the parameterized RayFlex Skid Buffer module and the ``defined-once-instantiated-everywhere" Shared RayFlex Data Structure. This makes RayFlex an ideal testing ground for academic research and exploration. Our implementation allows researchers to explore various design choices, fostering a realistic understanding of hardware ray tracer design trade-offs. Through comprehensive case studies, we demonstrate the versatility of RayFlex in evaluating different pipeline configurations and extending its functionality to support additional computational tasks. We show that by extending the functionality of a baseline RT unit datapath with an area cost of 36\% and a power overhead of about 20\%, the RT unit can calculate the Euclidean distance and cosine distance of vectors of arbitrary dimension, thereby accelerating a broader range of data-analytics workloads. The source code of RayFlex is available at \mbox{\url{https://github.com/purdue-aalp/rayflex}}.
\end{abstract}


% Hard-coded placement of textbox for IEEE copyright notice
\begin{textblock*}{16.8cm}(2.3cm,26cm) % {block width} (coords) 
\small© 2025 IEEE. Personal use of this material is permitted.  Permission from IEEE must be obtained for all other uses, in any current or future media, including reprinting/republishing this material for advertising or promotional purposes, creating new collective works, for resale or redistribution to servers or lists, or reuse of any copyrighted component of this work in other works.
\end{textblock*}

\section{Introduction}

In 3D computer graphics, ray tracing (RT) is a rendering technique that generates photorealistic scenes by simulating the behavior of light rays \cite{shirley2003rt_book}. The algorithm casts rays from the eye (camera) into a scene and calculates the intersections between the ray and objects; secondary rays are then cast from intersection points to model the effect of reflection, refraction, and shadows~\cite{whitted2005improved}. Ray tracing has been widely used in visual effects for still CGI images and movies for decades~\cite{techmonitor_art_1998}, but its high computational cost and divergent control flow in graphic processing units (GPUs)~\cite{lu2017unleashing} have restricted it to limited ray tracing effects in real-time rendering applications~\cite{wald2001interactive,PharrMatt2004PBRF} until the recent advent of real-time hardware RT units~\cite{amd_rdna2_2020,apple_m3_2023,beets_ray_2023,intel_xehpg_2022,nvidia_turing_2018}. Recent RT units improve the efficiency of ray tracing on GPUs by offloading intersection tests to fixed function datapaths, thus mitigating the SIMT divergence problem and freeing up general-purpose GPU cores for concurrent tasks~\cite{liu2021intersection,ha2024generalizing,hsu_paper_barnes}. 

The RT unit is programmed through graphics libraries like Vulkan, Optix, and DirectX~\cite{optix_paper,nvidia_optix,dxr_spec,vulkan_ray_tracing}. Programmers create an Acceleration Structure (AS) - typically a Bounding Volume Hierarchy (BVH) tree that groups triangle primitives of a scene into nested Axis-Aligned Bounding Boxes - and leverage the fixed function RT unit to efficiently traverse this Acceleration Structure to test for ray-box or ray-primitive intersections~\cite{saed2022vulkan,nah2011t,nah2014raycore,lee2013sgrt,schmittler2002saarcor,spjut2009trax,woop2005rpu}. Figure~\ref{fig:bvh-diagram} shows the structure of the BVH tree. However, ray tracing is not the only workload that the RT unit can accelerate: a variety of publications recently show that
many non-graphic problems can be cast to the ray tracing model and use the RT unit for acceleration~\cite{rtnn, rtindex, RadiusSearch,rt-force-directed,rt-mesh-location, rt-particle-track, bauer-visibility, arkade,nagarajan2023rtknns,nagarajan2023rtdbscan}. On the hardware side, recent work in GPU microarchitecture~\cite{hsu_paper_barnes,ha2024generalizing} proposed extending the hardware RT unit to accelerate nearest neighbor search and B-tree search in non-graphic workloads. This trend is reminiscent of the early days of general-purpose GPU programming when programmers made use of the graphics API to compute matrix multiplications~\cite{larsen2001fast} - what followed was an evolution of the GPU architecture and programming model which eventually enabled the GPU for general-purpose workloads. We envision a similar trend of generalization of the hardware RT unit. Therefore, an open model of the hardware RT unit would be beneficial and facilitate architecture research of the ray tracing unit.

Vulkan-Sim~\cite{saed2022vulkan} is currently the only public GPU architecture simulator that models detailed hardware ray tracing. Vulkan-Sim has a timing model that provides cycle-level simulation of the RT unit and focuses on the warp management and memory accesses of hardware ray tracing. However, for the low-level datapath that performs BVH operations (coordinate transformations, ray-box intersection tests, and ray-triangle intersection tests), Vulkan-Sim takes a black-box approach by simply assuming a fixed latency and omits internal details of the circuit. 

To shed light on the internals of the RT unit datapath and facilitate research in this area, we introduce RayFlex, a hardware RT unit datapath modeled at the register transfer level using Chisel~\cite{bachrach2012chisel}. It implements a fixed-latency pipeline that executes all aforementioned BVH operations, and the pipeline can be easily modified to support more operations, for instance, Euclidean and cosine distance calculation used by nearest neighbor search algorithms~\cite{hsu_paper_barnes}. 

This paper describes the architecture of RayFlex. RayFlex is designed to be simple and extensible:
(1) The IO interface is largely decoupled from the implementation of the datapath, allowing the user to explore different node configurations. For example, Rayflex can easily model a 4-wide BVH tree specified by the AMD RDNA2/3 Instruction Set Architectures (ISAs)~\cite{amd_rdna2_2020,amd_rdna3_2023_raytrace_insn} or a 6-wide BVH tree used in Mesa~\cite{ha2024generalizing}. 
(2) The functional unit (FU) pool in each stage can be modified separately so that users can study the trade-off between a shared FU pool design and a disjoint pipeline design or evaluate new operations beyond the ray-box or ray-triangle intersection tests.   
(3) RayFlex's datapath is designed as an elastic pipeline built from parameterized skid buffers modules, enabling the easy addition or deletion of pipeline stages and eliminating the need for a centralized controller.
(4) A Shared RayFlex Data Structure is defined once and instantiated for all pipeline stage registers, keeping the code complexity low.
(5) The precision and format of intermediate floating numbers are parameterized, enabling easy evaluation of different rounding strategies. 
(6) The module can be integrated with other open-source chip design frameworks such as Chipyard~\cite{amid2020chipyard}, the Vortex GPU~\cite{tine2021vortex}, or CGRA frameworks~\cite{tan2021aurora,weng2020dsagen}. 

Academic research in computer architecture often presents proof-of-concept designs highlighting their first-order gains, which calls for a low-maintenance, high-extensibility code base that facilitates rapid prototyping. RayFlex is designed with this in mind.  In Section~\ref{sec:rtl-design}, we introduce the parameterized RayFlex Skid Buffer module and a Shared RayFlex Data Structure. These two key design concepts contribute to RayFlex's extensibility and ease of development.

In Section~\hbox{\ref{sec:validation}}, we present our test cases for the functional correctness of RayFlex, we discuss the performance validation of RayFlex, and we highlight the complementary nature of RayFlex with Vulkan-Sim. Because the architecture of the hardware RT unit datapath used in commercial products is undisclosed, and because of the opaque software implementation of vendor-specific ray tracing pipelines, it isn't easy to correlate the throughput and initiation interval of RayFlex with RT units in real hardware. However, we reason that real GPUs likely contain hundreds of RT units equivalent in computing power to a RayFlex datapath.

To demonstrate the utility of RayFlex, in Section~\ref{sec:case-study} we present two case studies to evaluate (1) a unified pipeline for all BVH operations versus disjoint pipelines and (2) an extended RT unit that additionally supports the calculation of Euclidean and cosine distances. We present the evaluation methodology and results in Sections\mbox{~\ref{sec:methodology} and \ref{sec:eval}} before a brief discussion of related work in Section\mbox{~\ref{sec:related} and a conclusion in Section\mbox{~\ref{sec:conclusion}}}.

This paper makes the following contributions: 
\begin{itemize}
    \item A first-of-its-kind RTL implementation of the hardware RT unit datapath implemented in Chisel. Its extensible architecture allows researchers to study a variety of design choices for the RT unit datapath. 
    \item Two case studies evaluating the trade-offs of a unified pipeline and an extended RT unit datapath.
\end{itemize}

\section{Background}

\subsection{Ray Tracing}
\begin{figure}
    \centering
    \includegraphics[width=0.95\linewidth]{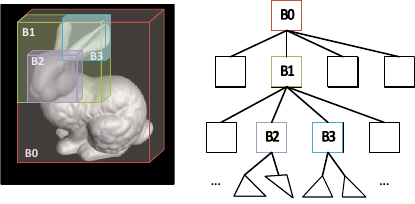}
    % \caption{sadfasdf}
        \caption{\textbf{Left}: The surface of this bunny consists of many triangle primitives. Triangles are grouped into nested Axis-Aligned Bounding Boxes to form a Hierarchical Bounding Volume (BVH). \textbf{Right}: Nested boxes become parents and children in the BVH tree.}
    \label{fig:bvh-diagram}
\end{figure}

GPUs render scenes from a collection of 3-dimensional primitives (typically triangles). To perform ray tracing, the GPU tests individual rays against the coordinates of primitives for intersection. Since each frame can be rendered from hundreds of thousands (if not millions) of primitives and millions of rays, it is imperative to perform this test efficiently. To this end, triangle primitives are grouped hierarchically into a tree of nested Axis-Aligned Bounding Boxes (AABB) so the scene can be searched hierarchically for any intersection. This tree is called the Bounding Volume Hierarchy (BVH), as illustrated in Figure~\ref{fig:bvh-diagram}. When traversing the BVH, ray-box intersection tests are performed on internal nodes and ray-triangle intersection tests are performed on leaf nodes. 

\subsection{RT Unit}
\begin{figure}
    \centering
    \includegraphics[width=1.0\linewidth]{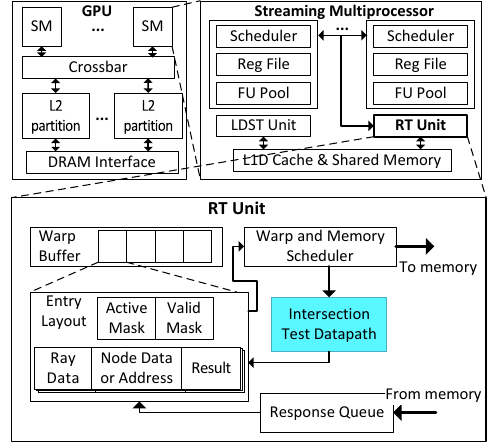}
    \caption{The GPGPU architecture and the RT Unit's position.}
    \label{fig:rt_unit_top}
\end{figure}

The RT Unit manages the traversal of the BVH by (1) scheduling memory requests to retrieve BVH node data and (2) using the RT unit datapath to test for intersections. The compute-intensive nature of intersection tests justifies the use of a fixed-function pipeline. 
An AMD patent\mbox{~\cite{saleh2021texture}} describes the methods and system for hardware ray tracing which has inspired the design of RayFlex.  Figure~\ref{fig:rt_unit_top} shows the high-level diagram of the RT unit, of which the highlighted structure is modeled in this paper. A detailed model of the RT unit can be found in Vulkan-Sim~\cite{saed2022vulkan}. A brief description of both the patent and Vulkan-Sim is given in Section\mbox{~\ref{sec:perf_validation}}.

\subsection{Intersection Tests} \label{sec:intersect-background}
The computation runtime of ray tracing is dominated by %retrieving BVH node data and 
ray-box and ray-triangle intersection tests.

\subsubsection{Ray-Box} \label{sec:ray-box-intersect}
Ray-box intersection tests determine which nodes to traverse to next in the hierarchy.
The slab method \cite{slab-pat, PharrMatt2004PBRF,slab-method, Majercik2018Voxel} is the most commonly used and prevalent algorithm to compute the intersection of a ray and AABB.
The general structure of the slab method is shown in Algorithm~\ref{alg:slab}.
For brevity, lines 1-4 show the computation along a single dimension, but it is important to note that each step is performed across the x, y, and z coordinates.
The inputs to the algorithm include a three-dimensional ray consisting of an origin and direction with a parametric time parameter $t$.
The slab method computes the intersection of a ray with the box by computing the intersection of the ray along each axis pair of planes.
An intersection occurs if and only if the intervals overlap.
The box is defined by a pair of three-dimensional points indicating their maximum and minimum corners.
We first translate these points to the ray origin using six addition operations (lines 1 and 2).
Next, the minimum and maximum t parameter values are computed for each of the three dimensions by dividing the box coordinate by the ray direction (lines 3 and 4).
A common optimization is to pre-compute the ray's inverse direction and instead use multiplication for this step (6 multiplications).
We are left with three intervals corresponding to the $t$ parameter values that the ray passes through each of the box's axis planes.
A fourth interval is provided as input to define the ray's extent, $t\_r\_beg$ and $t\_r\_end$.
The $tmin$ value (line 5) corresponds to the distance at which the ray enters the box and the $tmax$ value (line 6) corresponds to the distance the ray exits the box.
Therefore, line 7 returns the result of the intersection test and the distance at which the intersection occurred.
\begin{algorithm}
% \small
\caption{Slab Ray-box Intersection Method}\label{alg:slab}
    % \hspace*{\algorithmicindent} \textbf{Input} $ray\_origin$
    % \hspace*{\algorithmicindent} \textbf{Output} 
\begin{algorithmic}[1]
% \item[]/* Translate box to ray origin */

\STATE $box\_hi\gets box\_hi - ray\_origin$
\STATE $box\_lo\gets box\_lo - ray\_origin$
% \item[]
% \item[]/* Calculate intersection interval for each axis plane */
% \STATE $t\_min \gets box\_lo * ray\_inv$
% \STATE $t\_max \gets box\_hi * ray\_inv$
\STATE $t\_min \gets box\_lo \div ray\_dir$ (or $box\_lo * ray\_inv$)
\STATE $t\_max \gets  box\_hi \div ray\_dir$ (or $box\_hi * ray\_inv$)
% \item[]
% \item[]/* find max of the start of the 4 intervals */
\STATE $tmin \gets$ \bf{max}$(t\_min\_x, t\_min\_y, t\_min\_z, t\_r\_beg)$

% \item[]
% \item[]/* find min of the end of the 4 intervals */
\STATE $tmax \gets$ \bf{min}$(t\_max\_x, t\_max\_y,t\_max\_z, t\_r\_end)$

% \item[]
% \item[]/* Ray intersects iff $r\_min <= r\_max$ */
\STATE \bf{return} tuple($tmin <= tmax$, $tmin$)

\end{algorithmic}
\end{algorithm}

\subsubsection{Ray-Triangle} \label{sec:ray-triangle-intersect}
Ray-triangle intersection tests are performed at the leaf nodes of the BVH. To this end, the watertight method~\cite{Woop2013Watertight} achieves high accuracy and efficiency using single-precision floating-point operations. Here, we only provide a high-level summary of this algorithm since a full description is verbose and already given in Appendix A of Reference~\cite{Woop2013Watertight}. First, it renames the x, y, and z axes in a winding preserving way so that the greatest component of the ray direction is on the z-axis. It then performs an affine transformation on both the ray and the triangle such that the ray becomes the unit ray on the z-axis with origin $(0,0,0)$ and direction $(0,0,1)$. Finally, the barycentric coordinate of the ray's intersection with the triangle plane is calculated, from which the intersection distance can be derived and represented in the form of a numerator-denominator pair. It is worth noting that the renaming of axes and the affine transform matrix are the sole properties of the ray and require floating point divisions, therefore they can be pre-calculated at ray instantiation before BVH traversal.

\section{Functionality and Design} \label{sec:rtl-design}

\subsection{Connection to the Instruction Set Architecture}
\begin{figure}
    \centering
    \includegraphics[width=0.80\linewidth]{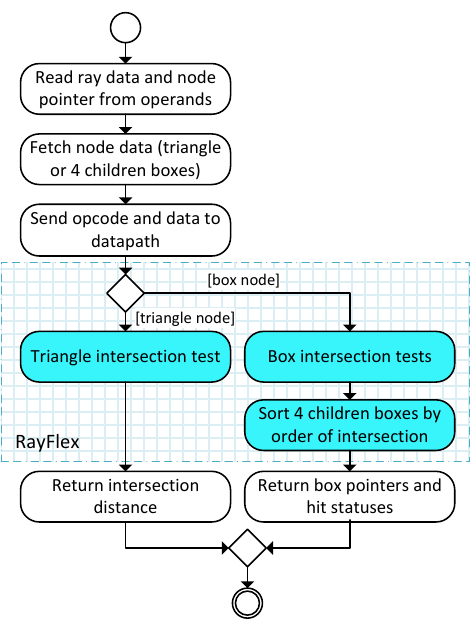}
    \caption{Activity diagram (per thread) of the \texttt{IMAGE\_BVH\_INTERSECT\_RAY} instruction. The highlighted activities happen inside RayFlex.}
    \label{fig:rdna3_bvh_intersect_insn}
\end{figure}

The IO specification of RayFlex is designed in reference to the latest RDNA3 ISA. RDNA3 defines a CISC-style SIMT \texttt{IMAGE\_BVH\_INTERSECT\_RAY} instruction~\cite{amd_rdna3_2023_raytrace_insn} for intersection tests. We summarize its activity in Figure~\ref{fig:rdna3_bvh_intersect_insn}. The ray data is directly passed by vector registers as operands of the instruction; the BVH node data is passed by a pointer. Depending on the node type, either one ray-triangle test or four ray-box tests are performed for each ray. Triangle tests return the intersection distance and triangle ID; box tests return the hit statuses and pointers to the four children boxes sorted by their order of intersection. In Figure~\ref{fig:rdna3_bvh_intersect_insn}, only the highlighted actions are performed by RayFlex. We follow the assumption that the management of warp threads is a duty of the higher-level RT unit~\cite{saed2022vulkan}, threads (rays) are individually fed to a single-lane datapath~\cite{hsu_paper_barnes,ha2024generalizing}, i.e., RayFlex does not execute in SIMD fashion (although jobs from individual threads can be pipelined).

\textbf{The IO specification} of RayFlex takes one opcode, one ray, one triangle, and four boxes as input. Each cycle, depending on the opcode, either the triangle or box data is valid.
Each box is defined by six floating point numbers (FPs) representing the minimum vertex and maximum vertex's coordinates. Each triangle is defined by nine FPs representing the three vertices' coordinates. Each ray is defined by three FPs for the origin point, three FPs for the direction vector, three FPs for the element-wise inverse of the direction vector, and one FP for the extent of the ray. As such, our ray format adheres to that specified by the RDNA3 ISA, but we additionally introduce six more FPs to the ray for the 3-dimensional \texttt{k} and \texttt{S} values to simplify the calculation done in RayFlex; these values correspond to the renaming of axes and the affine transform matrix which are pre-calculated at the time of ray creation as explained in Section~\ref{sec:ray-triangle-intersect}. The ISA does not specify the box and triangle formats, so we define our own as convenient. 

\subsection{BVH Operations}\label{sec:bvh_ops}
\subsubsection{Ray-Box Intersection}
During BVH traversal, the ray is tested for intersection against each of the children boxes of a box in which it has intersected. The RDNA3 ISA allows testing a ray against four children box nodes with a single instruction.

\begin{figure*}
    \centering
    \subfloat[][Four parallel ray-box intersection tests.]{\includegraphics[width=0.19\linewidth]{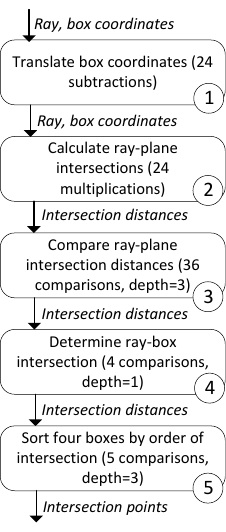}\label{fig:sub:ray-box-dataflow}}
  \hfill
  \subfloat[][Ray-triangle intersection test. Shaded steps are not performed in RayFlex.]{\includegraphics[width=0.41\linewidth]{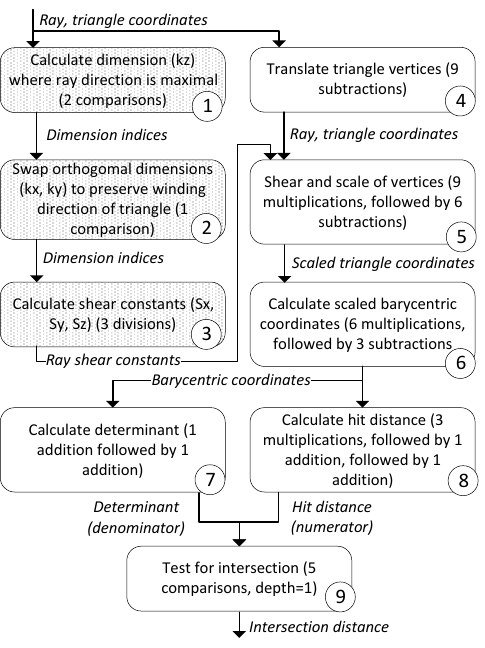}\label{fig:sub:ray-triangle-dataflow}}
  \hfill
  \subfloat[][Mapping from BVH operations to pipeline stages.]{\includegraphics[width=0.36\linewidth]{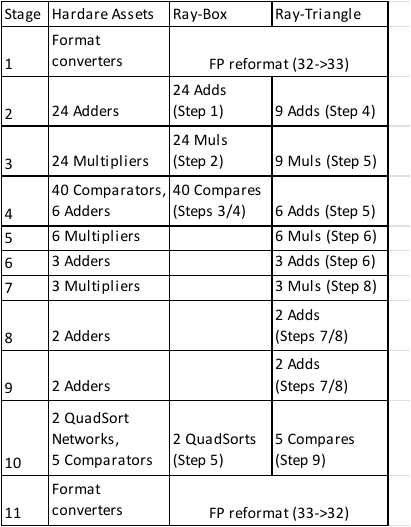}\label{fig:sub:fu_usage_baseline}}
  \caption{The dataflow and pipeline stages of RayFlex.} \label{fig:bvh_ops}
  %\description{Combination of old draft figures 4, 5, 8}
\end{figure*}

Figure~\ref{fig:sub:ray-box-dataflow} shows the high-level data flow of the \texttt{IMAGE\_BVH\_INTERSECT\_RAY} instruction when performing ray-box intersection tests. The first three steps \CircledText{1}\CircledText{2}\CircledText{3} perform four parallel ray-box tests, which we described in Section~\ref{sec:ray-box-intersect}. The last step \CircledText{4} sorts the four boxes by their order of intersection; a sorting network can finish this step using just five comparators~\cite{valsalam2013using}.

% \begin{figure}
%     \centering
%     \includegraphics[width=0.95\linewidth]{figs/fu_usage_baseline.pdf}
%     \caption{Mapping from BVH operations to pipeline stages.}
%     \label{fig:fu_usage_baseline}
% \end{figure}

% \begin{figure}
%     \centering
%     \includegraphics[width=0.4\linewidth]{figs/ray_box_dataflow.pdf}
%     \caption{The data flow of 4 parallel ray-box intersection tests.}
%     \label{fig:ray-box-dataflow}
% \end{figure}

\subsubsection{Ray-Triangle Intersection}
Figure~\ref{fig:sub:ray-triangle-dataflow} shows the high-level data flow of the watertight triangle intersection algorithm~\cite{Woop2013Watertight}. It can be broken down into the following steps: steps \CircledText{1}\CircledText{2} perform renaming of the axes so that the ray direction has its greatest component on the z-axis; step \CircledText{3} calculates the shear constants for the affine transform; steps \CircledText{4}\CircledText{5} apply the affine transform to the triangle; step \CircledText{6} calculates the barycentric coordinates of the intersection of the unit ray and the triangle plane; steps \CircledText{7}\CircledText{8}\CircledText{9} calculate the distance of intersection. 

The first three steps (shaded in gray) are pre-computed in the general-purpose GPU core at ray creation time. As explained at the end of Section\mbox{~\ref{sec:ray-triangle-intersect}}, the axis rotation and shear constants are the sole properties of the ray and involve divisions to calculate; therefore, it is sensible to treat such computation as part of the ray-instantiation routine and leverage the SIMD hardware of the GPU to perform these steps. Consequently, we avoid adding division units to RayFlex and reduce the pipeline depth.

% \begin{figure}
%     \centering
%     \includegraphics[width=0.8\linewidth]{figs/ray_triangle_dataflow.pdf}
%     \caption{The data flow of a ray-triangle intersection test. Shaded steps are not performed in RayFlex.}
%     \label{fig:ray-triangle-dataflow}
% \end{figure}

\subsection{Elastic Pipeline and Skid Buffers} \label{sec:skid_buffer}

The datapath of RayFlex adopts the \textit{elastic pipeline} architecture~\cite{cortadella2010elastic,vijayaraghavan2009bounded}. Pipeline stages use the two-phase bundled data convention~\cite{sutherland1989micropipelines} (aka. the ``valid-ready handshake protocol''~\cite{amba_axi_2021}) to transfer data and propagate back pressure. 

\begin{figure*}
    \centering
    \subfloat[][RTL diagram of a parameterized RayFlex Skid Buffer module with input data type T and output data type U.]{\includegraphics[width=0.43\linewidth]{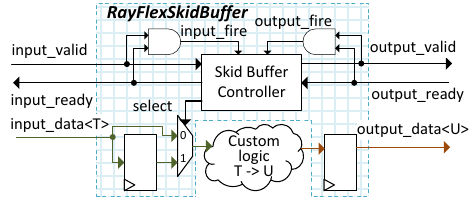}\label{fig:sub:skid_buffer_rtl}}
    \hfill
    \subfloat[][RayFlex's pipeline consists of a chain of RayFlex Skid Buffer modules. The first stage converts the input format to a wide  Shared RayFlex Data Structure (SRFDS), and the last stage converts the SRFDS to output format. All intermediate stages use the same
    SRFDS.]{\includegraphics[width=0.56\linewidth]{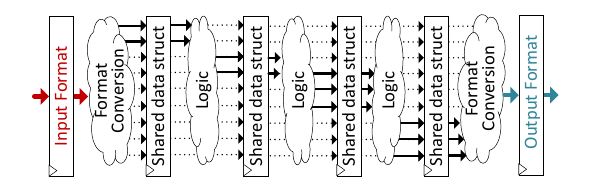}\label{fig:sub:global_data_bus}}
    \caption{The pipeline architecture of RayFlex.}
\end{figure*}

The building block of the pipeline of RayFlex is the RayFlex Skid Buffer module\mbox{~\cite{laforest2024skid}}, shown in Figure\mbox{~\ref{fig:sub:skid_buffer_rtl}}. It manages the synchronization with producer and consumer stages and encapsulates a chunk of programmer-supplied logic. We note that the programmer-supplied logic can be stateful, so one can use accumulators in the logic to sum the data from multiple beats. The RayFlex Skid Buffer module is parameterized by two data types, T and U, which represent the input and output data types of the programmer-supplied logic, respectively. Despite the parameterization, Chisel treats all RayFlex Skid Buffer modules like a single class. This makes programmatic handling of pipeline stages simple.

The self-synchronizing feature of the elastic pipeline eliminates the need for a global controller that dictates the stop-and-go of individual stages and decouples the control of pipeline stages from their functionality. Despite the higher area overhead of registers, we hope RayFlex's elastic pipeline makes it convenient for researchers to maintain and modify the datapath architecture.

% \begin{figure}
%     \centering
%     \subfloat[][RTL diagram of the RayFlex Skid Buffer module and its relation to user-supplied logic.]{\includegraphics[]{figs/skid_buffer_stage.pdf}\label{fig:skid_buffer_stage}}

%     \vfill
%     \subfloat[][Skid buffer controller finite state machine.]{\includegraphics[]{figs/skid_buffer_fsm.pdf}f}
   
%     \caption{Design of a RayFlex Skid Buffer.}
%     \label{fig:skid_buf}
% \end{figure}

\subsection{Unified Datapath Pipeline}

 RayFlex's pipeline has a throughput of 1 operation per cycle and a fixed latency of 11 cycles. 
 % Figure~\ref{fig:pipeline_arch} shows the high-level architecture of RayFlex (each skid buffer - stage logic pair corresponds to one instance of the circuit in Figure~\ref{fig:skid_buffer_stage}). 
 Figure~\ref{fig:sub:fu_usage_baseline} shows the mapping from BVH operation steps (annotated with circled numbers from Figures~\ref{fig:sub:ray-box-dataflow} and~\ref{fig:sub:ray-triangle-dataflow}) to pipeline stages. The logic of each stage is encapsulated in an instance of the RayFlex Skid Buffer module. For blank stages (e.g., stages 5 to 9 of the Ray-Box operation), the RayFlex Skid Buffer modules copy the input to output. Below, we highlight several design choices.
\begin{itemize}
    \item Steps that contain arithmetic operations that cannot be parallelized are spread to consecutive stages (e.g., step 8 of the ray-triangle operation is mapped to three stages);
    \item Steps that can be parallelized map to the same stage (e.g., steps 7 and 8 of the ray-triangle operation share the same stages);
    \item Steps that are sequential may be merged to one stage if doing so does not worsen the critical path (e.g., merging steps 3 and 4 of the ray-box operation to the same stage does not lengthen the critical path of RayFlex).
\end{itemize}

As reflected from the allocation of hardware assets, RayFlex implements a unified pipeline by sharing functional units at each stage between ray-box and ray-triangle operations. We will study an alternative design in the Case Study Section~\ref{sec:case_study_disjoint}, where we reserve private FUs for each operation. 

% \begin{figure}
%     \centering
%     \includegraphics[width=0.65\linewidth]{figs/pipeline_arch.pdf}
%     \caption{Top-level pipeline architecture of RayFlex.}
%     \label{fig:pipeline_arch}
% \end{figure}

% \begin{figure}
%     \centering
%     \includegraphics[width=0.95\linewidth]{figs/fu_usage_baseline.pdf}
%     \caption{Mapping from BVH operations to pipeline stages.}
%     \label{fig:fu_usage_baseline}
% \end{figure}

\subsection{Shared RayFlex Data Structure} \label{sec:global_data_struct}

% Benefit 1: define once, use everywhere, eliminate dead nodes as needed
% Benefit 2: complements the convenience of using elastic pipelines

In pipelined datapaths, the data registered at each stage is typically unique to that particular stage - at first sight, it is unnecessary and wasteful to keep a data field that was only used by the 1st stage in any subsequent stage's register or vice versa, reserve a data field that will not be produced until the last stage in any earlier stage's register.

However, we argue that manually specifying unique data structures for individual stages at the high level is a cumbersome process and undermines the extensibility and ease of maintenance of a research-oriented datapath like RayFlex. Instead, we define a very wide data structure containing all data fields that need to be registered at \textit{any} stage of the entire pipeline. We instantiate pipeline registers (skid buffers) of this same data structure everywhere through the pipeline (except at the first and last stage, where the IO specification has to be met). We call this data structure the ``\textit{Shared RayFlex Data Structure}'' of the pipeline.    

Figure~\ref{fig:sub:global_data_bus} illustrates the Shared RayFlex Data Structure running through a pipeline: except at the first and last stage where RayFlex performs the conversion between the internal Shared RayFlex Data Structure and the external IO data layout, the same data structure is used at all stages. The parameterized RayFlex Skid Buffer module introduced in Section\mbox{~\ref{sec:skid_buffer}} simplifies the creation of such a pipeline.

At the pre-synthesis RTL, it is legal for any stage to read any Shared RayFlex Data Structure field in its input and write to any Shared RayFlex Data Structure field in its output. However, a sane implementation of the stage logic refrains from reading and writing fields irrelevant to that stage; in return, unused wires and registers will eventually be identified as dead nodes and eliminated during the synthesis process. The result of eliminating dead nodes is manifested in the dashed arrows. 

When defining a stage logic at the Chisel level, we first directly assign the input Shared RayFlex Data Structure to the stage output register. After that, we may define custom logic to overwrite any data field that is supposed to be produced by this stage.   

The benefit of this Shared RayFlex Data Structure is two-fold. First, a single data structure is ``defined once and used everywhere,'' thus simplifying the pipeline design and lowering the labor overhead of adding new features or inserting new stages. Second, it complements the convenience of the elastic pipeline: the researcher can dedicate more time to the implementation of stage logic and reduce the time spent on managing the timing or interfaces of individual stages.   

% \begin{figure}
%     \centering
%     \includegraphics[width=1.0\linewidth]{figs/global_data_struct_hor.pdf}
%     \caption{At the RTL, a single Shared RayFlex Data Structure runs through all stages of the pipeline. Unused wires/registers (dashed arrows) will be eliminated during synthesis.}
%     \label{fig:global-data-struct}
% \end{figure}

\subsection{Floating Point Value Format and Rounding}
RayFlex processes single-precision floating point numbers. While the IO specification takes the standard FP32 format, internally, floating point values are represented by a special \textit{recorded format} that introduces an extra bit to the exponent field. This format simplifies the implementation of floating point circuits~\cite{hauser2022hardfloat}. We introduce a pair of extra stages at the input and output to convert between the standard FP32 and internal formats.

At each step of computation, the result of a full-precision arithmetic operation gains extra bits in both the exponent and significand. RayFlex performs rounding after every addition/multiplication. The rounding circuit is not trivial and adds to the overall area/power. 

While currently unexplored by this project, it is possible to forgo rounding at some or all stages to trade for a better area/frequency. We envision two potential challenges of this choice: 

    First, the design of a unified pipeline might become complicated when its functionality is extended to support new operations. Take, for example, the baseline ray-box and ray-triangle operations. At pipeline stage 6, the ray-box data has undergone one round of addition and one round of multiplication, whereas the ray-triangle data has undergone three additions and two multiplications. Their data will have different precision levels if no rounding was performed; if somehow we want to perform multiplication on the ray-box data at stage 7, then we need to align the precision of both data before sending them to the same functional unit. Although this is programmatically possible in Chisel~\cite{cook2017diplomatic}, doing so complicates the code structure. 
    
    Second, the final result produced by the hardware can deviate from the ``golden'' result produced by a software implementation due to the unusual rounding approach and, as a result, complicate verification.

\subsection{Summary}
We have presented the functionality and design of RayFlex. We introduced the IO specification and its connection to the RDNA3 intersection test instruction. We described the data flow of ray-box and ray-triangle intersection tests. We introduced the RayFlex Skid Buffer, the Shared RayFlex Data Structure, and the floating point formats of RayFlex.

We reiterate the two key design concepts of RayFlex. (1) The Shared RayFlex Data Structure leverages the logic synthesizers' dead node elimination feature. It allows the RTL designer to define-once and use-everywhere a single data structure for passing intermediate data through the pipeline. This makes RayFlex extensible. (2) The RayFlex Skid Buffer module, parameterized by two data types each representing its input and output data bundle, constitutes the entire pipeline. By encapsulating arbitrary logic blocks inside this same module class (notwithstanding the possibly different class parameters), the RTL designer can easily manage all pipeline stages of RayFlex programmatically.

\section{Validation} \label{sec:validation}

\subsection{Functional Correctness}
RayFlex is meant to be a realistic representation of the hardware RT unit datapath found in commercial hardware, it is therefore necessary to guarantee the functional correctness of RayFlex. To this end, we define twenty test cases to verify the functional correctness of RayFlex against a golden software implementation that serves as our ground truth.

Nine test cases target the ray-box intersection. We note that for cases where the ray is coplanar with one of the box surfaces, our hardware implementation and the gold reference treat them as misses. The reason is that at the numerical level if a component of a ray's direction vector is zero, positive or negative infinity may be used to represent the inverse of that value. When an FP functional unit multiplies infinity by zero (as in the case the ray is coplanar with a surface), the result is NaN, which returns \textit{false} for all $<,\le,=,\ge,>$ operations with any other number.

\textbf{Test cases for ray-box intersection:} (1) Ray originating from within the box (hit), (2) ray from outside the box and pointing away (miss), (3) ray from a surface of the box and pointing away (miss), (4) ray from a corner of the box and pointing away (miss), (5) ray from a corner of the box and pointing along an edge (miss), (6) ray from outside, pointing towards the box (hit), (7) ray hits two boxes in a row, (8) ray hits three boxes in a row and misses a fourth box off its path, (9) ray from outside the box, overlapping with an edge of the box (miss).
% \begin{enumerate}
%     \item \hl{Ray originating from within the box (hit).}
%     \item \hl{Ray from outside the box and pointing away (miss).}
%     \item \hl{Ray from a surface of the box and pointing away (miss).}
%     \item \hl{Ray from a corner of the box and pointing away (miss).}
%     \item \hl{Ray from a corner of the box and pointing along an edge (miss).}
%     \item \hl{Ray from outside, pointing towards the box (hit).}
%     \item \hl{Ray hits two boxes in a row.}
%     \item \hl{Ray hits three boxes in a row and misses a fourth box off its path.}
%     \item \hl{Ray from outside the box, overlapping with an edge of the box (miss).}
% \end{enumerate}

Eleven test cases target the ray-triangle intersection. Our implementation and gold reference adopt backface culling, which means the ray has to intersect the triangle on the ``front'' side to register a hit, i.e., a hit implies $\overrightarrow{r_{dir}}\cdot(\overrightarrow{AB}\times\overrightarrow{AC}) > 0$. Coplanar rays and triangles always miss. A non-coplanar ray intersecting the triangle at the edge or vertex is considered to be hitting the triangle.

\textbf{Test cases for ray-triangle intersection}: (1) ray hits the back of triangle (miss), (2) ray hits the front of triangle, (3) ray hits an edge of triangle from the front side (hit), (4) ray hits a triangle vertex from the front side (hit), (5) ray misses the triangle, (6) ray is parallel to the normal vector of the triangle but has no intersection (miss), (7) ray hits a far-away triangle, (8) ray hits the front of triangle at an oblique angle, (9) coplanar ray hits the edge of triangle (miss), (10) ray (aligned with a different axis compared to case \#2) hits the front of triangle, (11) coplanar ray originating from within the triangle hits edge of triangle (miss).
% \begin{enumerate}
%     \item \hl{Ray hits the back of triangle (miss).}
%     \item \label{item:ray_hit_front_triangle}\hl{Ray hits the front of triangle.}
%     \item \hl{Ray hits an edge of triangle from the front side (hit).}
%     \item \hl{Ray hits a triangle vertex from the front side (hit).}
%     \item \hl{Ray misses the triangle.}
%     \item \hl{Ray is parallel to the normal vector of the triangle but has no intersection (miss).}
%     \item \hl{Ray hits a far-away triangle.}
%     \item \hl{Ray hits the front of triangle at an oblique angle.}
%     \item \hl{Coplanar ray hits the edge of triangle (miss).}
%     \item \hl{Ray (aligned with a different axis compared to case \mbox{\ref{item:ray_hit_front_triangle}}) hits the front of triangle.}
%     \item \hl{Coplanar ray originating from within the triangle hits edge of triangle (miss).}
% \end{enumerate}

\subsection{RayFlex Performance Relative to Hardware and Simulation} \label{sec:perf_validation}
Functional correctness alone does not mean RayFlex is a realistic representation of what real hardware does in commercial products; highly correlated performance metrics are a more convincing proof. We have nonetheless refrained from performing a quantitative performance validation of RayFlex for the two reasons described in this section.

First, we could not find any public report explicitly disclosing the architecture of the hardware RT unit datapath, it would therefore be misleading to claim RayFlex as any equivalent representation or substitute of its counterpart in commercial products. An AMD patent\mbox{~\cite{saleh2021texture}} inspired our work on RayFlex; while this patent specifies how the BVH tree is traversed and how ray and node data from active threads in a wave (also known as a warp, which is typically a collection of 32 or 64 threads) are serialized into individual ``transactions'' tested for intersection in a MIMD (multiple instruction multiple data) fashion, the patent does not claim exactly how intersection tests are performed for each transaction. It specifically declares that the tests could be performed by various kinds of processors, e.g., general purpose processors, digital signal processors, ASICs, FPGAs, etc., and/or a state machine. RayFlex should be considered one manifestation of the processor for the intersection tests. 

Second, it is challenging to measure the throughput and initiation interval of the hardware RT unit datapath in real hardware, not only because it is not easy to program the RT unit at the lowest level directly but also because of the opaque implementation of vendor-specific ray tracing pipelines and data structures. For example, the BVH structure is not publicly documented and we have reason to believe they are more complex than what is shown in Figure\mbox{~\ref{fig:bvh-diagram}} to support more complex features like motion-blur\mbox{~\cite{NVIDIA_AMPERE_GA102_GPU_ARCHITECTURE}} and micro-meshes\mbox{~\cite{NVIDIA_ADA_GPU_ARCHITECTURE}}.

However, we can still perform a quick check to compare the throughput of RayFlex with real RT units. The NVIDIA Turing GPU is the first commercial product to include a hardware RT unit and it can yield 100 tera-ops of compute per second for ray-tracing\mbox{~\cite{nvidia_turing_2018}}. To calculate how many operations each RT unit can perform every cycle, we look at the Quadro RTX 6000 card and find it contains 72 RT units and runs at a clock speed of 1455 MHz. This means every unit can perform $100e12 / 72 / 1455e6 \approx 955$ operations per cycle. Figure\mbox{~\ref{fig:sub:fu_usage_baseline}} indicates that the RayFlex pipeline can perform a maximum of 125 operations per cycle (we optimistically assume all functional units are active and treat each multiplier, adder, and comparator as contributing one operation per cycle, and we treat each QuadSort unit as containing five comparators; we do not consider the computation done in the format converters of the first and last stage). This result indicates that each of the 72 RT units in the Quadro RTX 6000 card likely contains at least $955 / 125 \approx 7.6$ independent processors equivalent in computing power to a RayFlex datapath. Given these high-level numbers, a designer can leverage RayFlex to model a full RT unit with equivalent throughput through a combination of clocking the unit at a higher rate and instantiating several parallel RayFlex pipelines. When implementing a full RT unit the number of parallel datapaths is a tunable design decision, as indicated in AMD's patented RT unit design\mbox{~\cite{saleh2021texture}}.

RayFlex is complementary to Vulkan-Sim\mbox{~\cite{saed2022vulkan}}, the high-level simulator for ray tracing workloads. Vulkan-Sim provides an open implementation of the Vulkan ray tracing pipeline (commonly kept in obscurity by GPU manufacturers) and defines a performance model of the RT unit. The whole RT unit includes the fixed-function datapath for intersection tests and the warp management and memory scheduling logic necessary for the recursive traversal of the BVH tree. Despite providing an extensive model, Vulkan-Sim treats the intersection test datapath as an opaque functional unit whose latency and throughput values can be arbitrarily defined. In contrast, RayFlex zooms in on the implementation details of the datapath (module highlighted in Fig\mbox{~\ref{fig:rt_unit_top}}). Researchers interested in studying the ray tracer datapath can use RayFlex as the testing ground to determine how much hardware is required to realize certain functionalities and the latencies/bandwidths.

The original Vulkan-Sim paper\mbox{~\cite{saed2022vulkan}} models the functional unit for intersection tests as described by Liu et al. \mbox{~\cite{liu2021intersection}} which assumed a 2-cycle latency. Vulkan-Sim instantiates sufficient instances of the unit to avoid the need for adding another queue, so their RT functional unit pool's initiation interval is likely no worse than one ray per cycle. We conclude that the configurations used by Vulkan-Sim are more optimistic than those indicated by RayFlex.

\section{Case Studies} \label{sec:case-study}

We present two case studies using RayFlex to extend the datapath for new operations and examine an alternative disjoint pipelines architecture.

\subsection{Adding New Operations}
There have been many reports of people successfully using the hardware RT unit to accelerate non-graphics workloads~\cite{rtnn, rtindex, RadiusSearch,rt-force-directed,rt-mesh-location, rt-particle-track, bauer-visibility, arkade,nagarajan2023rtknns,nagarajan2023rtdbscan}, it is therefore interesting to consider extending the functionality of the hardware RT unit so that it can accelerate such workloads more efficiently. One of the most common types of program that the hardware RT unit can accelerate is the hierarchical search: the searched dataset is typically represented as a collection of spheres in the 3-dimensional space and hierarchically grouped into a BVH tree, and the query point is expressed as a ray in this space with a tiny extent. A search hit is indicated by the intersection of the query ray with any sphere. 

\begin{figure*}
    \centering
    \subfloat[][Euclidean distance.]{\includegraphics[width=0.19\linewidth]{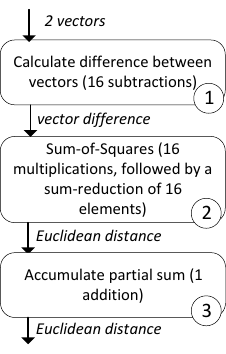}\label{fig:sub:euclidean-dataflow}}
  \hfill
  \subfloat[][Cosine distance.]{\includegraphics[width=0.38\linewidth]{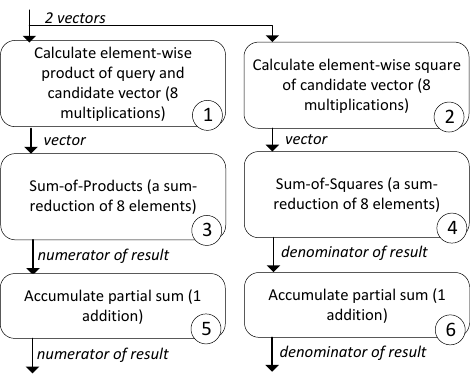}\label{fig:sub:cosine-dataflow}}
  \hfill
  \subfloat[][Mapping from new operations to pipeline stages.]{\includegraphics[width=0.42\linewidth]{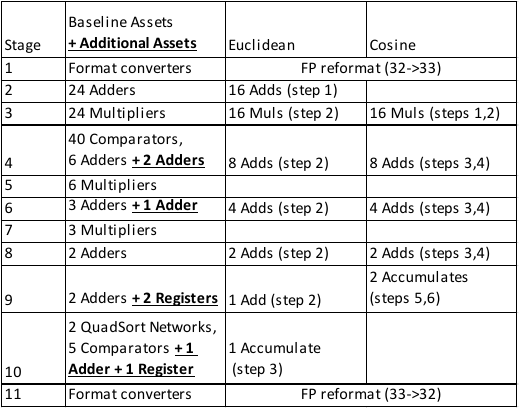}\label{fig:sub:fu_usage_newOps}}
  \caption{The dataflow and pipeline stages of two new operations.} \label{fig:new_ops}
  %\description{Combination of old draft figures 10, 11, 12}
\end{figure*}

To improve the utility of the hardware RT unit in such non-graphics workloads, Barnes et al.~\cite{hsu_paper_barnes} propose extending the RT unit datapath to calculate the Euclidean distance (Figure~\ref{fig:sub:euclidean-dataflow}), cosine distance (aka. angular distance, Figure~\ref{fig:sub:cosine-dataflow}), and key comparison in arbitrary dimensional space. They note the high overlap of functional units involved in these operations with the existing ray-box/ray-triangle intersection tests: just a few more adders, multipliers, and accumulators are necessary to support the new operations. 

In this case study, we evaluate such a feature-enhanced pipeline by adding the necessary functional units for both the Euclidean and cosine distance calculation and evaluate the power and area overhead in Section~\ref{sec:eval_new_ops}. The mapping of the data flow graph of new operations to the pipeline stages is shown in Figure~\ref{fig:sub:fu_usage_newOps}. 

\textbf{The IO specification} of the extended pipeline builds on the IO specification of the original RayFlex. Without touching the fields for the ray-box/ray-triangle operations, we add the following input fields: two sixteen-element vectors of FP32 values \texttt{euclidean\_a} and \texttt{euclidean\_b}, one sixteen-element bitmask \texttt{euclidean\_mask}, one boolean value \texttt{reset\_accumulator}; we add the following output fields: an FP32 value \texttt{euclidean\_accumulator}, a boolean value \texttt{euclidean\_reset}, an FP32 value \texttt{angular\_dot\_product}, an FP 32 value \texttt{angular\_norm}, and a boolean value \texttt{angular\_reset}.

% \begin{enumerate}
%     \item \hl{input [0:15] euclidean\_a [31:0]}
%     \item \hl{input [0:15] euclidean\_b [31:0]}
%     \item \hl{input [0:15] euclidean\_mask [0:0]}
%     \item \hl{input reset\_accumulator [0:0]}
%     \item \hl{output euclidean\_accumulator [31:0]}
%     \item \hl{output euclidean\_reset [0:0]}
%     \item \hl{output angular\_dot\_product [31:0]}
%     \item \hl{output angular\_norm [31:0]}
%     \item \hl{output augular\_reset [0:0]}
% \end{enumerate}

Among the inputs, \texttt{euclidean\_a} and \texttt{b} are each 16 FP32 values representing the Euclidean coordinates of points in the sixteen-dimensional space, but RayFlex can process points of arbitrarily high dimension by digesting their values over multiple beats: the boolean \texttt{reset\_accumulator} should be set on the last beat of a pair of very-high dimension vectors. Finally, the \texttt{euclidean\_mask} is a bit mask that can invalidate any dimension of \texttt{euclidean\_a} and \texttt{b}. 

Among the outputs, \texttt{euclidean\_accumulator} is an FP32 value for the squared value of the Euclidean distance between the two input vectors. \texttt{angular\_dot\_product} and \texttt{angular\_norm} are the two value (FP32) calculated for accelerating cosine distances. The two boolean signals \texttt{euclidean\_reset} and \texttt{angular\_reset} correspond to the input \texttt{reset\_accumulator} signal from eleven cycles ago, which indicates that the current beat of the output is the last beat of a pair of very-high dimension vectors.

We note that \texttt{reset\_accumulator} is the only synchronous signal that can clear the accumulator registers for Euclidean or cosine distance operations. Consequently, a pair of very long vectors that take multiple cycles to transmit to the RayFlex for Euclidean or cosine distance calculation can be interspersed by any amount of ray-box or ray-triangle operations so long as the \texttt{reset\_accumulator} signal is not set until the last beat of the vector data. Likewise, multi-cycle Euclidean and multi-cycle cosine jobs can intersperse each other because they use separate accumulators.
% \begin{figure}
%     \centering
%     \includegraphics[width=0.45\linewidth]{figs/euclidean_dataflow.pdf}
%     \caption{The data flow of arbitrary-dimension Euclidean distance calculation.}
%     \label{fig:euclidean-dataflow}
% \end{figure}

% \begin{figure}
%     \centering
%     \includegraphics[width=0.9\linewidth]{figs/angular_dataflow.pdf}
%     \caption{The data flow of arbitrary-dimension cosine distance calculation.}
%     \label{fig:cosine-dataflow}
% \end{figure}

% \begin{figure}
%     \centering
%     \includegraphics[width=0.9\linewidth]{figs/fu_usage_new_ops.pdf}
%     \caption{Mapping from new operations to pipeline stages.}
%     \label{fig:fu_usage_newOps}
% \end{figure}

Ha et al.~\cite {ha2024generalizing} likewise propose adding point-to-point distance calculation to the RT unit's functionality and note their minimal changes to the existing datapath. 

\subsection{Disjoint Pipelines} \label{sec:case_study_disjoint}

The implementation details of the hardware RT unit datapath in commercial products are anything but public, so any proposal to extend the datapath has to make assumptions on the baseline architecture. This paper follows the design choice of RayCore~\cite{raycore} and HSU~\cite{hsu_paper_barnes}, which use a unified pipeline for both BVH operations to improve functional unit reuse. In contrast, TTA~\cite{ha2024generalizing} assumes \textit{disjoint} (i.e., separate) pipelines, so each operation mode has its own pool of functional units. We therefore use RayFlex to evaluate an alternative design where ray-box and ray-triangle data still enter the same pipeline, but the two operations use private FUs at each stage. This means the datapath needs to provision more FUs. In addition, we investigate the joint effects of adding new operations to a disjoint-pipeline design. We present the results in Section~\ref{sec:eval_disjoint}.

\section{Methodology} \label{sec:methodology}

We implement RayFlex with Chisel (version 5.0.0) and develop test benches using the chiseltest library~\cite{chiseltest2024}. The RTL design is verified at the behavioral level with special cases and hundreds of thousands of random test cases, covering all ray-box, ray-triangle, Euclidean, and cosine operations described in Sections~\ref{sec:rtl-design} and~\ref{sec:case-study}. 

RayFlex sources floating point functional units from the Berkeley Hardfloat library~\cite{berkeley_hardfloat_repo,hauser2022hardfloat}.

We use a 15nm process design kit~\cite{martins2015open} to synthesize the design in the Cadence Genus Synthesis tool (ver. 21.11). Unless otherwise specified, the design synthesis is performed with a clock of 1GHz. 
The area and power information are extracted from Genus reports. 
We use VCD-format stimulus files each collected from a real testbench of 100 random test cases to calculate power. 

This paper explores a design space of three dimensions:
\begin{enumerate}
    \item Various target clock frequencies (from 500MHz to 1500MHz);
    \item A ``baseline'' pipeline that supports only ray-box and ray-triangle operations, and an ``extended'' pipeline that additionally supports Euclidean and cosine distance calculation~\cite{hsu_paper_barnes};
    \item A ``unified'' pipeline in which functional units (FUs) are shared among operations in each stage, and a ``disjoint'' pipelines design in which each operation uses its own FUs (however, all operations enter the same pipeline).
\end{enumerate}

\section{Evaluation} \label{sec:eval}

This section examines the overhead of incorporating new functionality into the datapath in both the unified pipeline and the disjoint pipelines configuration.   

\subsection{Area} \label{sec:eval_new_ops}

\begin{figure*}
    \centering
    \includegraphics[width=\linewidth]{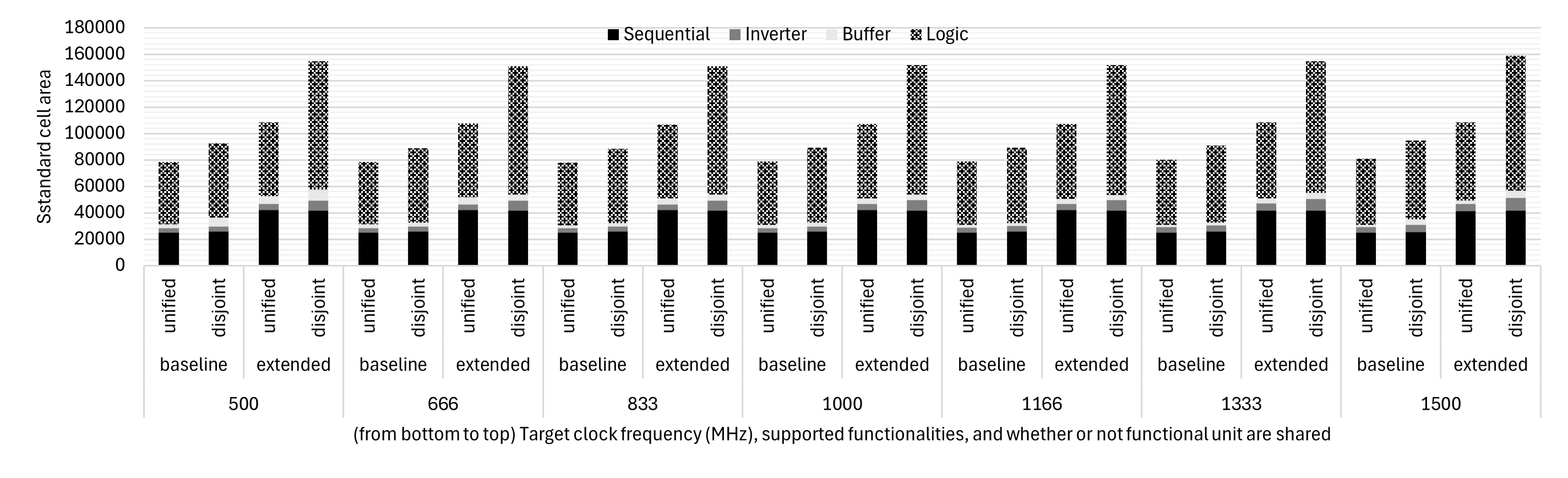}
    \caption{Circuit area versus (1) various target clock frequencies, (2) whether or not the pipeline supports new operations (``baseline'' vs ``extended''), and (3) whether or not functional units are shared at each stage (``unified'' vs ``disjoint'').}
    \label{fig:area_v_frequency}
\end{figure*}

Figure~\ref{fig:area_v_frequency} shows the circuit area when synthesized at various clock frequencies. Most notably, the circuit area does not show much sensitivity to the target clock frequency in our evaluated range. The baseline-unified design has the lowest area; making it ``disjoint'' adds about 13\% overhead, extending it with support for Euclidean and cosine distance calculations adds about 36\% overhead; finally, making it both ``disjoint'' and ``extended'' adds about 92\% overhead, or 70\% overhead compared with the baseline-disjoint design.

The circuit area is decomposed into four categories: sequential, inverter, buffer, and logic. The sequential and logic components dominate the total area in all cases, so, we focus on these two components. 

We first examine the consequence of moving from the unified pipeline design to disjoint pipelines, with all other variables fixed. The ``sequential'' area stays almost constant, but the ``logic'' area increases by about 18\% and 74\% for ``baseline'' and ``extended'', respectively. This increase in ``logic'' area is primarily due to the provisioning of more private FUs.

We then study the effects of adding new functionalities (going from ``baseline'' to ``extended''). This results in an increase in the ``sequential'' and ``logic'' areas. (1) The fluctuations in the ``logic'' area show sensitivity to the FU sharing strategy: the ``logic'' area grows by just around 17\% in the shared pipeline but grows by 72\% in the disjoint pipelines. This unsurprising observation reflects the compounding benefit of introducing new operations to the unified datapath. The more operations the shared pipeline can support, the more area-efficient it is. (2)  The ``sequential'' area surprisingly grows by about 64\% regardless of the functional unit-sharing strategy. Our explanation for this consistently high overhead is that we designed RayFlex to use disjoint pipeline registers for each operation regardless of the FU-sharing strategy. As a result (despite the dead node elimination performed during synthesis), at each stage during any active cycle, only a portion of stage registers contain valid data for the active operation, and other registers hold invalid data. 

We have deliberately chosen to store intermediate data in disjoint pipeline registers in favor of a clean and simple architecture, resulting in a high overhead of the ``sequential'' area when new operations are added. Alternatively, it is possible to share pipeline registers among operations. This can be done by using the \texttt{.asTypeOf} method on the Shared RayFlex Data Structure (described in Section~\ref{sec:global_data_struct}) to cast it into the correct data layout for distinct operation modes (similar to accessing a \textit{union} in C/C++). Additionally, this alternative design unlocks an interesting opportunity for optimization: Since ``dead'' bits of the Shared RayFlex Data Structure are eliminated during synthesis, the layout pattern of individual data structures for each operation should be optimally aligned to maximize the amount of dead node elimination. The goal is to map fields with the same lifetime to the same position in the Shared RayFlex Data Structure. On the other hand, the worst case occurs when every bit of the Shared RayFlex Data Structure remains live in all stages due to any operation; in this case, dead node elimination cannot eliminate any bit of the Shared RayFlex Data Structure.  

\subsection{Power Consumption} \label{sec:eval_disjoint}

\begin{figure}
    \centering
    \includegraphics[width=0.95\linewidth]{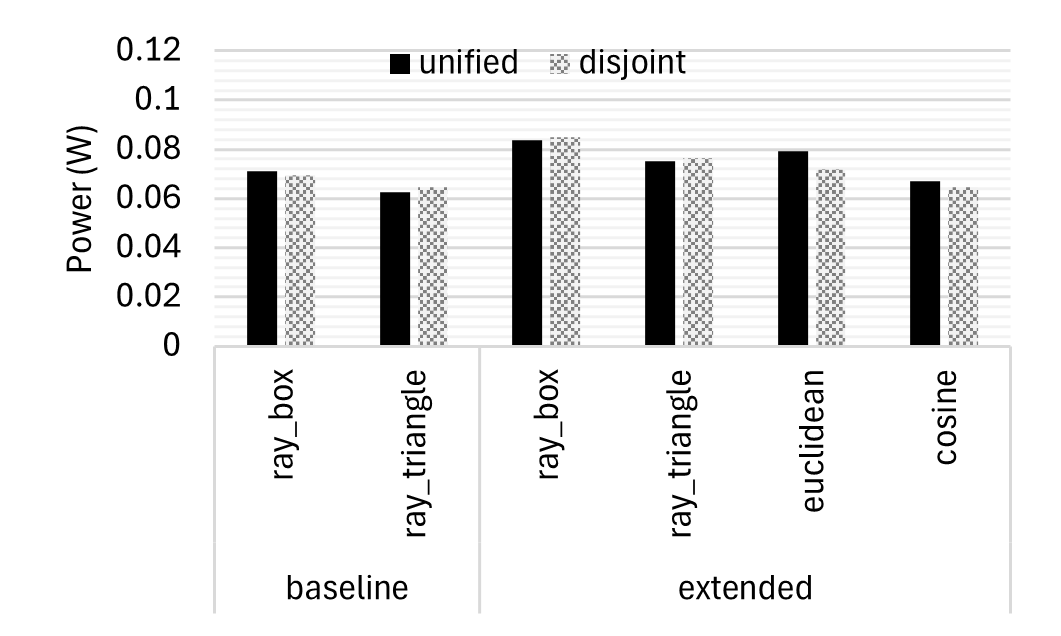}
    \caption{Power consumption when executing different operations at full throughput.}
    \label{fig:power_v_op}
\end{figure}

Figure~\ref{fig:power_v_op} shows the total power consumption of RayFlex when processing data for each operating mode at full throughput. The power numbers are measured for RayFlex synthesized with a target clock frequency of 1000MHz. In all cases, the power is between 60mW and 85mW.

We first examine the effect of adding new operations (Euclidean and cosine distance) to the unified pipeline. Compared with the baseline, the extended datapath increases the power consumption of the ray-box and ray-triangle intersection tests by 18\% and 20\%, respectively. This is the power overhead of the additional pipeline stage registers introduced to support new operations.

We then examine the effect of using private FUs for each operation, i.e., we take the baseline ``unified'' pipeline and make it ``disjoint''. For ray-box and ray-triangle operations, the change in power is within +/-2.5\% which we consider minuscule. For Euclidean and cosine operations, however, the power consumption reduces by 9\% and 3\% respectively.

Why do ray-box and ray-triangle operations see minimal change in power consumption despite the introduction of additional FUs? This is because RayFlex gates the input port of each FU (mostly adders and multipliers) with a multiplexer which always feeds zero to the FU unless the operation needs the output of this FU. This minimizes the dynamic power consumption of RayFlex. With the technology library\mbox{~\cite{martins2015open}} we use, the amount of static power is an order of magnitude smaller than the amount of dynamic power, therefore, the power overhead of adding more FUs is minimal.

What is more surprising is the 9\% decrease in power consumption by Euclidean operations in the ``disjoint'' design. A stage-by-stage breakdown of power consumption shows that all power savings come from stage 3 where RayFlex uses multipliers to calculate the element-wise product of vectors; the fluctuation of power in all other stages remains minimal. After careful analysis and hypothesis testing, we conclude that this decrease of power in stage 3 is the result of the logic synthesizer specializing some of the multipliers (\texttt{y=a*b}) into squarers (\texttt{y=a*a}) which have lower area and power cost\mbox{~\cite{moore2013low}}. The Euclidean operation performs 16 squares at this stage, and the cosine operation performs 8 squares and 8 general multiplications; the disjoint pipelines design therefore gives the synthesizer the opportunity to specialize and simplify some multipliers into squarers. Indeed, once we intentionally perturb the logic in stage 3 of the disjoint pipelines design such that none of the multipliers receive both inputs from the same wires, this decrease in power consumption disappears, and the power of Euclidean operations increases by 1.9\% compared with the baseline.

\subsection{Power Sensitivity to Target Clock Frequency} 
\begin{figure}
    \centering
    \includegraphics[width=0.97\linewidth]{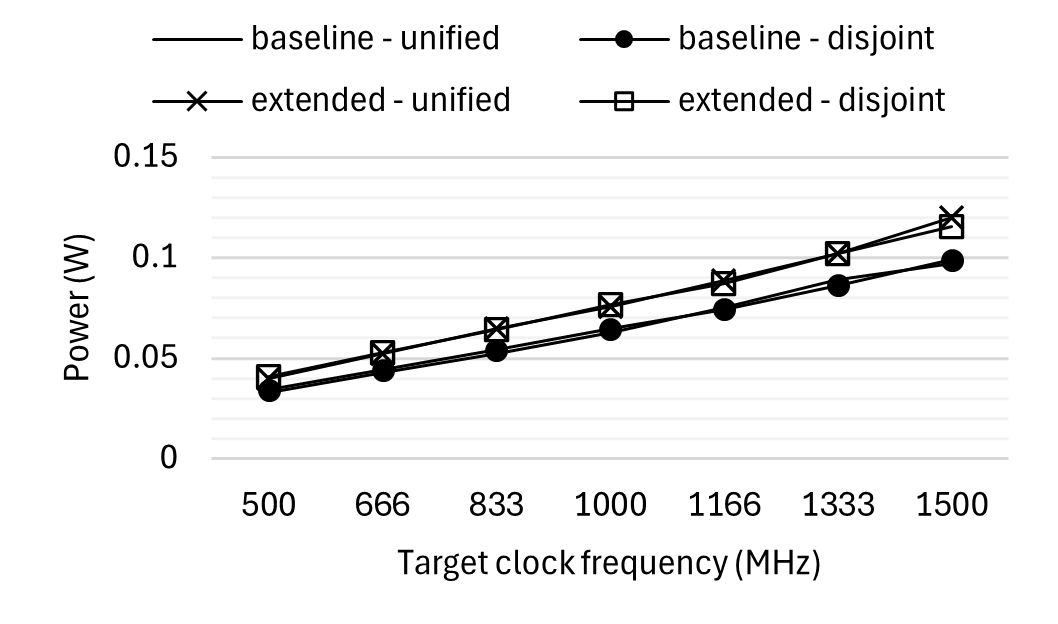}
    \caption{Power consumption of ray-triangle operations when RayFlex is synthesized at various target clock frequencies.}
    \label{fig:power_v_freq}
\end{figure}

Figure~\ref{fig:power_v_freq} focuses on the power consumption of RayFlex running ray-triangle operations when the circuit is synthesized at various target clock frequencies. The power consumption follows a nearly linear trend in the evaluated frequency range. The power difference between ``unified'' and ``disjoint'' remains small (within the range of +/-4\%) across the frequency range, and the power difference between ``baseline'' and ``extended'' varies between 14\% and 22\%.

\section{Related work} \label{sec:related}

\textbf{Simulators for RT. }
Vulkan-Sim~\cite{saed2022vulkan} studies the functional and timing behavior of the hardware RT unit in an execution-driven approach by interfacing with the Vulkan library. It focuses on the warp management and memory accesses
involved in hardware ray tracing but does not model the internals of the RT unit datapath. Zatel~\cite{grigoryan2024zatel} reduces the simulation time of Vulkan-Sim 
by means of sampling.

\textbf{RT-accelerated applications. }
Beyond graphics workloads, a number of works have leveraged the RT unit for acceleration. 
RTNN~\cite{rtnn}, RT-kNNS Unbound~\cite{nagarajan2023rtknns}, Arkade~\cite{arkade}, RT-DBSCAN~\cite{nagarajan2023rtdbscan}, and Fast Radius Search~\cite{RadiusSearch} reformulate the nearest neighbor search problem to match the ray tracing programming interface. 
RTIndeX~\cite{rtindex} reformulates the indexing of the database with the ray tracing model. 
Reference~\cite{rt-force-directed} accelerates force-directed graph drawing.
Reference~\cite{rt-mesh-location} accelerates the calculation of mesh point locations in unstructured volume rendering. 
Reference~\cite{rt-particle-track} uses hardware RT units to accelerate a particle tracking program.  
Reference~\cite{bauer-visibility} reduces the problem of dependence analysis and coherence of implicitly parallel programming systems to ray tracing.

\textbf{Extended hardware RT units. } HSU~\cite{hsu_paper_barnes} and TTA~\cite{ha2024generalizing} propose extending the hardware RT unit datapath with minimal changes in exchange for better programmability and applicability to a broader range of workloads (e.g., tree traversal).  TTA+\cite{ha2024generalizing} decomposes the fixed function RT unit datapath into a pool of individual programmable operation units. 

\section{Conclusion} \label{sec:conclusion}
A wide variety of workloads can be accelerated by the hardware RT unit in general-purpose GPUs. Recently, we have seen several proposals extending the hardware RT unit for better programmability and utility to a broader range of workloads. We feel the need to release an open-source RTL design of the RT unit datapath to shed light on the internals of this relatively novel module and facilitate research in this area. Hence, we introduce RayFlex, a first-of-its-kind open-source RTL implementation of the RT unit datapath. We explain that the architecture of RayFlex is designed to be extensible and friendly to researchers interested in studying the datapath. We present two case studies where we extend the datapath with support for calculating Euclidean and cosine distance and compare it against an alternative ``disjoint pipelines'' design. As we detail our design choices and evaluate our case studies, we shed light on some interesting alternative design choices that can be explored by future work, for example, floating point rounding strategies and shared pipeline stage registers. 

\section*{Availability}
The source code of RayFlex is hosted on GitHub: \mbox{\url{https://github.com/purdue-aalp/rayflex}}.

\section*{Acknowledgment}
We thank our reviewers and shepherd for helping us improve this paper. This work was supported, in part, by NSF CCF \#1943379 (CAREER).

\bibliographystyle{IEEEtran}
\bibliography{bib}

\begin{thebibliography}{10}
\providecommand{\url}[1]{#1}
\csname url@samestyle\endcsname
\providecommand{\newblock}{\relax}
\providecommand{\bibinfo}[2]{#2}
\providecommand{\BIBentrySTDinterwordspacing}{\spaceskip=0pt\relax}
\providecommand{\BIBentryALTinterwordstretchfactor}{4}
\providecommand{\BIBentryALTinterwordspacing}{\spaceskip=\fontdimen2\font plus
\BIBentryALTinterwordstretchfactor\fontdimen3\font minus
  \fontdimen4\font\relax}
\providecommand{\BIBforeignlanguage}[2]{{%
\expandafter\ifx\csname l@#1\endcsname\relax
\typeout{** WARNING: IEEEtran.bst: No hyphenation pattern has been}%
\typeout{** loaded for the language `#1'. Using the pattern for}%
\typeout{** the default language instead.}%
\else
\language=\csname l@#1\endcsname
\fi
#2}}
\providecommand{\BIBdecl}{\relax}
\BIBdecl

\bibitem{shirley2003rt_book}
P.~P. Shirley and R.~K. Morley, \emph{\BIBforeignlanguage{eng}{Realistic ray
  tracing}}, 2nd~ed.\hskip 1em plus 0.5em minus 0.4em\relax Natick, Mass: AK
  Peters, 2003.

\bibitem{whitted2005improved}
T.~Whitted, ``An improved illumination model for shaded display,'' in \emph{ACM
  Siggraph 2005 Courses}, 2005, pp. 4--es.

\bibitem{techmonitor_art_1998}
{TechMonitor}. (1998) Advanced rendering technology has new 3d graphics chip.
  TechMonitor.

\bibitem{lu2017unleashing}
Y.~L{\"u}, L.~Huang, L.~Shen, and Z.~Wang, ``Unleashing the power of gpu for
  physically-based rendering via dynamic ray shuffling,'' in \emph{Proceedings
  of the 50th Annual IEEE/ACM International Symposium on Microarchitecture},
  2017, pp. 560--573.

\bibitem{wald2001interactive}
I.~Wald, P.~Slusallek, C.~Benthin, and M.~Wagner, ``Interactive rendering with
  coherent ray tracing,'' in \emph{Computer graphics forum}, vol.~20,
  no.~3.\hskip 1em plus 0.5em minus 0.4em\relax Wiley Online Library, 2001, pp.
  153--165.

\bibitem{PharrMatt2004PBRF}
M.~Pharr, G.~Humphreys, and P.~Hanrahan,
  \emph{\BIBforeignlanguage{eng}{Physically Based Rendering: From Theory to
  Implementation}}, 2nd~ed.\hskip 1em plus 0.5em minus 0.4em\relax Chantilly:
  Elsevier Science \& Technology, 2004.

\bibitem{amd_rdna2_2020}
{Advanced Micro Devices, Inc.}, ``{RDNA 2 Instruction Set Architecture},''
  Advanced Micro Devices, Inc., Santa Clara, CA, Technical Documentation,
  November 2020.

\bibitem{apple_m3_2023}
{Apple Inc.} (2023, October) Apple unveils {M3}, {M3 Pro}, and {M3 Max}, the
  most advanced chips for a personal computer. Apple Inc.

\bibitem{beets_ray_2023}
K.~Beets, ``Ray tracing for the masses,'' Imagination Technologies, White
  Paper, January 2023.

\bibitem{intel_xehpg_2022}
{Intel Corporation}, ``Introduction to the {Xe-HPG} architecture,'' Intel
  Corporation, White Paper, 2022, document describing Intel Arc A-Series GPUs
  architecture.

\bibitem{nvidia_turing_2018}
{NVIDIA Corporation}, ``{NVIDIA Turing GPU Architecture: Graphics
  Reinvented},'' NVIDIA Corporation, White Paper WP-09183-001\_v01, 2018.

\bibitem{liu2021intersection}
L.~Liu, W.~Chang, F.~Demoullin, Y.~H. Chou, M.~Saed, D.~Pankratz, T.~Nowicki,
  and T.~M. Aamodt, ``Intersection prediction for accelerated gpu ray
  tracing,'' in \emph{MICRO-54: 54th Annual IEEE/ACM International Symposium on
  Microarchitecture}, 2021, pp. 709--723.

\bibitem{ha2024generalizing}
D.~Ha, L.~Liu, Y.~H. Chou, S.~Go, W.~W. Ro, H.-W. Tseng, and T.~M. Aamodt,
  ``Generalizing ray tracing accelerators for tree traversals on gpus,'' in
  \emph{2024 57th IEEE/ACM International Symposium on Microarchitecture
  (MICRO)}.\hskip 1em plus 0.5em minus 0.4em\relax IEEE, 2024, pp. 1041--1057.

\bibitem{hsu_paper_barnes}
A.~Barnes, F.~Shen, and T.~G. Rogers, ``Extending gpu ray-tracing units for
  hierarchical search acceleration,'' in \emph{57th IEEE/ACM International
  Symposium on Microarchitecture (MICRO)}, 2024.

\bibitem{optix_paper}
\BIBentryALTinterwordspacing
S.~G. Parker, J.~Bigler, A.~Dietrich, H.~Friedrich, J.~Hoberock, D.~Luebke,
  D.~McAllister, M.~McGuire, K.~Morley, A.~Robison, and M.~Stich, ``Optix: a
  general purpose ray tracing engine,'' \emph{ACM Trans. Graph.}, vol.~29,
  no.~4, Jul. 2010. [Online]. Available:
  \url{https://doi.org/10.1145/1778765.1778803}
\BIBentrySTDinterwordspacing

\bibitem{nvidia_optix}
{NVIDIA Corporation}, ``Nvidia optix 8.0 documentation,''
  \url{https://raytracing-docs.nvidia.com/optix8/index.html}, 2024, accessed:
  2024-12-13.

\bibitem{dxr_spec}
{Microsoft Corporation}, ``Directx raytracing (dxr) functional spec,''
  \url{https://microsoft.github.io/DirectX-Specs/d3d/Raytracing.html}, 2024,
  accessed: 2024-12-13.

\bibitem{vulkan_ray_tracing}
{Khronos Group}, ``Ray tracing in vulkan,''
  \url{https://www.khronos.org/blog/ray-tracing-in-vulkan}, 2024, accessed:
  2024-12-13.

\bibitem{saed2022vulkan}
M.~Saed, Y.~H. Chou, L.~Liu, T.~Nowicki, and T.~M. Aamodt, ``Vulkan-sim: A gpu
  architecture simulator for ray tracing,'' in \emph{2022 55th IEEE/ACM
  International Symposium on Microarchitecture (MICRO)}.\hskip 1em plus 0.5em
  minus 0.4em\relax IEEE, 2022, pp. 263--281.

\bibitem{nah2011t}
J.-H. Nah, J.-S. Park, C.~Park, J.-W. Kim, Y.-H. Jung, W.-C. Park, and T.-D.
  Han, ``T\&i engine: Traversal and intersection engine for hardware
  accelerated ray tracing,'' in \emph{Proceedings of the 2011 SIGGRAPH Asia
  Conference}, 2011, pp. 1--10.

\bibitem{nah2014raycore}
J.-H. Nah, H.-J. Kwon, D.-S. Kim, C.-H. Jeong, J.~Park, T.-D. Han, D.~Manocha,
  and W.-C. Park, ``Raycore: A ray-tracing hardware architecture for mobile
  devices,'' \emph{ACM Transactions on Graphics (TOG)}, vol.~33, no.~5, pp.
  1--15, 2014.

\bibitem{lee2013sgrt}
W.-J. Lee, Y.~Shin, J.~Lee, J.-W. Kim, J.-H. Nah, S.~Jung, S.~Lee, H.-S. Park,
  and T.-D. Han, ``Sgrt: A mobile gpu architecture for real-time ray tracing,''
  in \emph{Proceedings of the 5th high-performance graphics conference}, 2013,
  pp. 109--119.

\bibitem{schmittler2002saarcor}
J.~Schmittler, I.~Wald, and P.~Slusallek, ``Saarcor: a hardware architecture
  for ray tracing,'' in \emph{Proceedings of the ACM SIGGRAPH/EUROGRAPHICS
  conference on Graphics hardware}, 2002, pp. 27--36.

\bibitem{spjut2009trax}
J.~Spjut, A.~Kensler, D.~Kopta, and E.~Brunvand, ``Trax: A multicore hardware
  architecture for real-time ray tracing,'' \emph{IEEE Transactions on
  Computer-Aided Design of Integrated Circuits and Systems}, vol.~28, no.~12,
  pp. 1802--1815, 2009.

\bibitem{woop2005rpu}
S.~Woop, J.~Schmittler, and P.~Slusallek, ``Rpu: a programmable ray processing
  unit for realtime ray tracing,'' \emph{ACM Transactions on Graphics (TOG)},
  vol.~24, no.~3, pp. 434--444, 2005.

\bibitem{rtnn}
\BIBentryALTinterwordspacing
Y.~Zhu, ``{RTNN: Accelerating Neighbor Search Using Hardware Ray
  Tracing}.''\hskip 1em plus 0.5em minus 0.4em\relax New York, NY, USA:
  Association for Computing Machinery, 2022, p. 76–89. [Online]. Available:
  \url{https://doi.org/10.1145/3503221.3508409}
\BIBentrySTDinterwordspacing

\bibitem{rtindex}
\BIBentryALTinterwordspacing
J.~Henneberg and F.~Schuhknecht, ``{RTIndeX: Exploiting Hardware-Accelerated
  GPU Raytracing for Database Indexing},'' \emph{Proceedings of the VLDB
  Endowment}, vol.~16, no.~13, p. 4268–4281, 2023. [Online]. Available:
  \url{https://doi.org/10.14778/3625054.3625063}
\BIBentrySTDinterwordspacing

\bibitem{RadiusSearch}
\BIBentryALTinterwordspacing
I.~Evangelou, G.~Papaioannou, K.~Vardis, and A.~A. Vasilakis, ``{Fast Radius
  Search Exploiting Ray Tracing Frameworks},'' \emph{Journal of Computer
  Graphics Techniques (JCGT)}, vol.~10, no.~1, pp. 25--48, February 2021.
  [Online]. Available: \url{http://jcgt.org/published/0010/01/02/}
\BIBentrySTDinterwordspacing

\bibitem{rt-force-directed}
S.~Zellmann, M.~Weier, and I.~Wald, ``{Accelerating Force-Directed Graph
  Drawing with RT Cores},'' in \emph{2020 IEEE Visualization Conference (VIS)},
  2020, pp. 96--100.

\bibitem{rt-mesh-location}
N.~Morrical, I.~Wald, W.~Usher, and V.~Pascucci, ``{Accelerating Unstructured
  Mesh Point Location With RT Cores},'' \emph{IEEE Transactions on
  Visualization and Computer Graphics}, vol.~28, no.~8, pp. 2852--2866, 2022.

\bibitem{rt-particle-track}
B.~Wang, I.~Wald, N.~Morrical, W.~Usher, L.~Mu, K.~Thompson, and R.~Hughes,
  ``{An GPU-accelerated particle tracking method for Eulerian–Lagrangian
  simulations using hardware ray tracing cores},'' \emph{Computer Physics
  Communications}, vol. 271, p. 108221, 2022.

\bibitem{bauer-visibility}
M.~Bauer, E.~Slaughter, S.~Treichler, W.~Lee, M.~Garland, and A.~Aiken,
  ``{Visibility Algorithms for Dynamic Dependence Analysis and Distributed
  Coherence}.''\hskip 1em plus 0.5em minus 0.4em\relax New York, NY, USA:
  Association for Computing Machinery, 2023, p. 218–231.

\bibitem{arkade}
D.~K. Mandarapu, V.~Nagarajan, A.~Pelenitsyn, and M.~Kulkarni, ``{Arkade:
  k-Nearest Neighbor Search With Non-Euclidean Distances using GPU Ray
  Tracing},'' in \emph{Proceedings of the 38th ACM International Conference on
  Supercomputing}.\hskip 1em plus 0.5em minus 0.4em\relax Association for
  Computing Machinery, 2024, p. 14–25.

\bibitem{nagarajan2023rtknns}
\BIBentryALTinterwordspacing
V.~Nagarajan, D.~Mandarapu, and M.~Kulkarni, ``Rt-knns unbound: Using rt cores
  to accelerate unrestricted neighbor search,'' in \emph{Proceedings of the
  37th ACM International Conference on Supercomputing}, ser. ICS '23.\hskip 1em
  plus 0.5em minus 0.4em\relax New York, NY, USA: Association for Computing
  Machinery, 2023, p. 289–300. [Online]. Available:
  \url{https://doi.org/10.1145/3577193.3593738}
\BIBentrySTDinterwordspacing

\bibitem{nagarajan2023rtdbscan}
V.~Nagarajan and M.~Kulkarni, ``Rt-dbscan: Accelerating dbscan using ray
  tracing hardware,'' in \emph{2023 IEEE International Parallel and Distributed
  Processing Symposium (IPDPS)}, 2023, pp. 963--973.

\bibitem{larsen2001fast}
E.~S. Larsen and D.~McAllister, ``Fast matrix multiplies using graphics
  hardware,'' in \emph{Proceedings of the 2001 ACM/IEEE Conference on
  Supercomputing}, 2001, pp. 55--55.

\bibitem{bachrach2012chisel}
\BIBentryALTinterwordspacing
Bachrach, Richards, Lee, Waterman, Avizienis, Wawrzynek, and Asanovic,
  ``Chisel: Constructing hardware in a scala embedded language,'' \emph{Design
  Automation Conference}, p. 1212–1221, Jan. 2012. [Online]. Available:
  \url{http://yadda.icm.edu.pl/yadda/element/bwmeta1.element.ieee-000006241660}
\BIBentrySTDinterwordspacing

\bibitem{amd_rdna3_2023_raytrace_insn}
{Advanced Micro Devices, Inc.}, ``{RDNA 3 Instruction Set Architecture},''
  Advanced Micro Devices, Inc., Santa Clara, CA, Technical Documentation,
  February 2023.

\bibitem{amid2020chipyard}
A.~Amid, D.~Biancolin, A.~Gonzalez, D.~Grubb, S.~Karandikar, H.~Liew,
  A.~Magyar, H.~Mao, A.~Ou, N.~Pemberton \emph{et~al.}, ``Chipyard: Integrated
  design, simulation, and implementation framework for custom socs,''
  \emph{IEEE Micro}, vol.~40, no.~4, pp. 10--21, 2020.

\bibitem{tine2021vortex}
B.~Tine, K.~P. Yalamarthy, F.~Elsabbagh, and K.~Hyesoon, ``Vortex: Extending
  the risc-v isa for gpgpu and 3d-graphics,'' in \emph{MICRO-54: 54th Annual
  IEEE/ACM International Symposium on Microarchitecture}, 2021, pp. 754--766.

\bibitem{tan2021aurora}
C.~Tan, C.~Xie, A.~Li, K.~J. Barker, and A.~Tumeo, ``Aurora: Automated
  refinement of coarse-grained reconfigurable accelerators,'' in \emph{2021
  Design, Automation \& Test in Europe Conference \& Exhibition (DATE)}.\hskip
  1em plus 0.5em minus 0.4em\relax IEEE, 2021, pp. 1388--1393.

\bibitem{weng2020dsagen}
J.~Weng, S.~Liu, V.~Dadu, Z.~Wang, P.~Shah, and T.~Nowatzki, ``Dsagen:
  Synthesizing programmable spatial accelerators,'' in \emph{2020 ACM/IEEE 47th
  Annual International Symposium on Computer Architecture (ISCA)}.\hskip 1em
  plus 0.5em minus 0.4em\relax IEEE, 2020, pp. 268--281.

\bibitem{saleh2021texture}
S.~J. Saleh, M.~V. Kazakov, and V.~Goel, ``Texture processor based ray tracing
  acceleration method and system,'' Dec.~14 2021, uS Patent 11,200,724.

\bibitem{slab-pat}
T.~T. Karras, T.~O. Alia, S.~M. Laine, and J.~E. Lindholm, ``{Beam Tracing},''
  U.S. Patent US10242485B2 Mar. 26, 2019.

\bibitem{slab-method}
T.~L. Kay and J.~T. Kajiya, ``{Ray Tracing Complex Scenes},'' in
  \emph{Proceedings of the 13th Annual Conference on Computer Graphics and
  Interactive Techniques (SIGGRAPH)}.\hskip 1em plus 0.5em minus 0.4em\relax
  Association for Computing Machinery, 1986, p. 269–278.

\bibitem{Majercik2018Voxel}
A.~Majercik, C.~Crassin, P.~Shirley, and M.~McGuire, ``{A Ray-Box Intersection
  Algorithm and Efficient Dynamic Voxel Rendering},'' \emph{Journal of Computer
  Graphics Techniques (JCGT)}, vol.~7, no.~3, pp. 66--81, 2018.

\bibitem{Woop2013Watertight}
\BIBentryALTinterwordspacing
S.~Woop, C.~Benthin, and I.~Wald, ``Watertight ray/triangle intersection,''
  \emph{Journal of Computer Graphics Techniques (JCGT)}, vol.~2, no.~1, pp.
  65--82, June 2013. [Online]. Available:
  \url{http://jcgt.org/published/0002/01/05/}
\BIBentrySTDinterwordspacing

\bibitem{valsalam2013using}
V.~K. Valsalam and R.~Miikkulainen, ``Using symmetry and evolutionary search to
  minimize sorting networks,'' \emph{The Journal of Machine Learning Research},
  vol.~14, no.~1, pp. 303--331, 2013.

\bibitem{cortadella2010elastic}
J.~Cortadella, M.~Galceran-Oms, and M.~Kishinevsky, ``Elastic systems,'' in
  \emph{Eighth ACM/IEEE International Conference on Formal Methods and Models
  for Codesign (MEMOCODE 2010)}.\hskip 1em plus 0.5em minus 0.4em\relax IEEE,
  2010, pp. 149--158.

\bibitem{vijayaraghavan2009bounded}
M.~Vijayaraghavan \emph{et~al.}, ``Bounded dataflow networks and
  latency-insensitive circuits,'' in \emph{2009 7th IEEE/ACM International
  Conference on Formal Methods and Models for Co-Design}.\hskip 1em plus 0.5em
  minus 0.4em\relax IEEE, 2009, pp. 171--180.

\bibitem{sutherland1989micropipelines}
I.~E. Sutherland, ``Micropipelines,'' \emph{Communications of the ACM},
  vol.~32, no.~6, pp. 720--738, 1989.

\bibitem{amba_axi_2021}
{Arm Limited}, ``{AMBA AXI and ACE Protocol Specification},'' Arm Limited,
  Technical Documentation IHI 0022Q, 2021, aRM ID: 102202.

\bibitem{laforest2024skid}
\BIBentryALTinterwordspacing
C.~E. LaForest, ``Pipeline skid buffer,'' in \emph{FPGA Design Elements}, 2024,
  ch. Pipeline Skid Buffer. [Online]. Available:
  \url{https://fpgacpu.ca/fpga/Pipeline_Skid_Buffer.html}
\BIBentrySTDinterwordspacing

\bibitem{hauser2022hardfloat}
\BIBentryALTinterwordspacing
J.~Hauser, ``{Berkeley Hardfloat},'' 2022. [Online]. Available:
  \url{http://www.jhauser.us/arithmetic/HardFloat.html}
\BIBentrySTDinterwordspacing

\bibitem{cook2017diplomatic}
H.~Cook, W.~Terpstra, and Y.~Lee, ``Diplomatic design patterns: A tilelink case
  study,'' in \emph{1st Workshop on Computer Architecture Research with
  RISC-V}, vol.~23, 2017.

\bibitem{NVIDIA_AMPERE_GA102_GPU_ARCHITECTURE}
\BIBentryALTinterwordspacing
``{Nvidia Ampere GA102 GPU Architecture},'' Nvidia Corporation. [Online].
  Available:
  \url{https://images.nvidia.com/aem-dam/en-zz/Solutions/geforce/ampere/pdf/NVIDIA-ampere-GA102-GPU-Architecture-Whitepaper-V1.pdf}
\BIBentrySTDinterwordspacing

\bibitem{NVIDIA_ADA_GPU_ARCHITECTURE}
\BIBentryALTinterwordspacing
``{Nvidia ADA GPU Architecture},'' Nvidia Corporation. [Online]. Available:
  \url{https://images.nvidia.com/aem-dam/Solutions/geforce/ada/nvidia-ada-gpu-architecture.pdf}
\BIBentrySTDinterwordspacing

\bibitem{raycore}
\BIBentryALTinterwordspacing
J.-H. Nah, H.-J. Kwon, D.-S. Kim, C.-H. Jeong, J.~Park, T.-D. Han, D.~Manocha,
  and W.-C. Park, ``Raycore: A ray-tracing hardware architecture for mobile
  devices,'' \emph{ACM Trans. Graph.}, vol.~33, no.~5, sep 2014. [Online].
  Available: \url{https://doi.org/10.1145/2629634}
\BIBentrySTDinterwordspacing

\bibitem{chiseltest2024}
\BIBentryALTinterwordspacing
{UC Berkeley Architecture Research}, ``chiseltest: The batteries-included
  testing and formal verification library for {Chisel}-based {RTL} designs,''
  2024, archived Project. [Online]. Available:
  \url{https://github.com/ucb-bar/chiseltest}
\BIBentrySTDinterwordspacing

\bibitem{berkeley_hardfloat_repo}
``{Berkeley Hardfloat},'' \url{https://github.com/ucb-bar/berkeley-hardfloat},
  UC Berkeley Architecture Research, 2023.

\bibitem{martins2015open}
M.~Martins, J.~M. Matos, R.~P. Ribas, A.~Reis, G.~Schlinker, L.~Rech, and
  J.~Michelsen, ``Open cell library in 15nm freepdk technology,'' in
  \emph{Proceedings of the 2015 Symposium on International Symposium on
  Physical Design}, 2015, pp. 171--178.

\bibitem{moore2013low}
J.~Moore, M.~A. Thornton, and D.~W. Matula, ``Low power floating-point
  multiplication and squaring units with shared circuitry,'' in \emph{2013 IEEE
  56th International Midwest Symposium on Circuits and Systems (MWSCAS)}.\hskip
  1em plus 0.5em minus 0.4em\relax IEEE, 2013, pp. 1395--1398.

\bibitem{grigoryan2024zatel}
D.~Grigoryan, Y.~H. Chou, and T.~M. Aamodt, ``Zatel: Sample complexity-aware
  scale-model simulation for ray tracing,'' in \emph{2024 IEEE International
  Symposium on Performance Analysis of Systems and Software (ISPASS)}.\hskip
  1em plus 0.5em minus 0.4em\relax IEEE, 2024, pp. 156--166.

\end{thebibliography}


% \end{IEEEbiography}


% % insert where needed to balance the two columns on the last page with
% %\newpage



%\vfill


\end{document}